\documentclass[12pt,preprint]{aastex}

\slugcomment{submitted to ApJ}

\shorttitle{GRMHD Simulation of Jet Formation}
\shortauthors{Nishikawa et al.}

\begin{document}

\title{A General Relativistic Magnetohydrodynamic Simulation of
Jet Formation}

\author{K.-I. Nishikawa\altaffilmark{1}}
\affil{National Space Science and Technology Center,
  320 Sparkman Drive, SD 50, Huntsville, AL 35805}
\email{ken-ichi.nishikawa@nsstc.nasa.gov}
\author{G. Richardson\altaffilmark{1}}
\affil{Department of Mechanical and Aerospace Engineering
University of Alabama in Huntsville
Huntsville, AL 35899}
\author{S. Koide}
\affil{Faculty of Engineering, Toyama University,
  3190 Gofuku, Toyama 930-8555 Japan}
\author{K. Shibata}
\affil{Kwasan and Hida Observatories, Kyoto University,
  Yamashina, Kyoto 607-8417 Japan}
\author{T. Kudoh}
\affil{Department of Physics and Astronomy,
   University of Western Ontario,
   London, Ontario N6A 3K7, Canada}
\author{P. Hardee}
\affil{Department of Physics and Astronomy,
  The University of Alabama,
  Tuscaloosa, AL 35487}

\and

\author{G. J. Fishman}
\affil{NASA-Marshall Space Flight Center,
  National Space Science and Technology Center,
  320 Sparkman Drive, SD 50, Huntsville, AL 35805}

\altaffiltext{1}{NRC Associate / NASA Marshall Space Flight Center}

\begin{abstract}

We have performed a fully three-dimensional general
relativistic magnetohydrodynamic (GRMHD) simulation of jet formation
from a thin accretion disk around a Schwarzschild black hole with a
free-falling corona. The initial simulation results show that a bipolar
jet (velocity $\sim 0.3c$) is created as shown by previous
two-dimensional axisymmetric simulations with mirror symmetry at the
equator.  The 3-D simulation ran over one hundred light-crossing time
units ($\tau_{\rm S} = r_{\rm S}/c$ where $r_{\rm S} \equiv 2GM/c^2$)
which is considerably longer than the previous simulations. We show
that the jet is initially formed as predicted due in part to magnetic
pressure from the twisting the initially uniform magnetic field and
from gas pressure associated with shock formation in the region around
$r = 3 r_{\rm S}$.  At later times, the accretion disk becomes thick
and the jet fades resulting in a wind that is ejected from the surface
of the thickened (torus-like) disk. It should be noted that no
streaming matter from a donor is included at the outer boundary in the
simulation (an isolated black hole not binary black hole). The wind
flows outwards with a wider angle than the initial jet. The widening of
the jet is consistent with the outward moving torsional Alfv\'{e}n waves
(TAWs). This
evolution of disk-jet coupling suggests that  the jet fades with
a thickened accretion disk due to the lack of streaming material from
an accompanying star.

\end{abstract}


\section{Introduction}

Relativistic jets have been observed in active galactic nuclei (AGNs)
(e.g. Urry \& Pavovani 1995; Ferrari 1998; Blandford 2002), in
microquasars in our Galaxy (e.g. Mirabel \& Rodriguez 1999), and it is
believed that they originate in the regions near accreting (stellar)
black holes and neutron stars \citep{mku01}.  To investigate the
dynamics of accretion disks and the associated jet formation, we have
performed jet formation simulations using a full 3-D GRMHD code.
This magnetic-acceleration mechanism has been proposed not only for AGN
jets, but also for (nonrelativistic) protostellar jets (see
\citet{mku01}). Kudoh, Matsumoto, \& Shibata (1998) found that the
terminal velocity of a jet is comparable to the rotational velocity of
the disk at the foot of the jet in nonrelativistic MHD simulations, and
that a relativistic jet should be formed near the event horizon. It
should be noted that the mechanism of jet generation in neutron star
binaries may be different from black hole binaries.  Recently, an
ultra-relativistic outflow was observed from an accreting stellar-mass
neutron star (Fender et al. 2004). Fender et al. (2004) have concluded that the
generation of highly relativistic outflows does not require properties
unique to black holes, such as an event horizon. It seems plausible
that production of ultrarelativistic jets may be a generic feature of
accretion onto neutron stars such as Cir-X and Sco X-1 at or near to
the Eddington limit. This result differs from the suggestions that
the velocities of jets would be limited to the surface escape speed
(Livio 1999).

The fluorescent iron line (K$_{\alpha}$) seen in the X-ray spectrum of
many active galactic nuclei (AGNs), and particularly in Seyfert
galaxies, has provided us with an extremely useful probe of the region
close to an accreting black hole. This is an important diagnostic with
which to study Doppler and gravitational redshifts, thereby providing
key information on the local kinematics of the material (see the
reviews by Fabian et al. 2000 and Reynolds \& Nowak 2003). The
simultaneous X-ray/IR flares from a black hole/relativistic jet system
are the observational evidence revealing a time-dependent interaction
between the jet and inner disk (e.g. Ueda et al. 2002). In regions
where disk-jet interactions take place near a black hole, the plasma
and magnetic fields interact in a complicated manner;
therefore a 3-dimensional GRMHD simulation such as we perform here is
required to investigate dynamics near
the horizon and ultimately to couple predicted dynamics to the iron line.

\citet{ksk99} have investigated in 2-D, the dynamics of an accretion
disk initially threaded by a uniform magnetic field in a
non-rotating corona (either in a state of steady fall or in
hydrostatic equilibrium) around a non-rotating (Schwarzschild) black
hole. The numerical results show that matter in the disk loses
angular momentum by magnetic braking, then falls into the black
hole.  The disk falls faster in this simulation than in the
non-relativistic case because of general-relativistic effects that
are important below $3 r_{\rm S}$, where $r_{\rm S} \equiv 2GM/c^2$
is the Schwarzschild radius. A centrifugal barrier at $r = 2 r_{\rm
S}$ strongly decelerates the infalling material. Plasma near the
shock at the centrifugal barrier is accelerated by the ${\bf J}
\times {\bf B}$ force and forms bipolar jets. Inside this {\it
magnetically driven jet}, the gradient of gas pressure also
generates a jet above the shock region ({\it gas-pressure driven
jet}). This {\it two-layered jet structure} is formed both in a
hydrostatic corona and in a steady-state falling corona. Koide et
al. (2000) have also developed a new GRMHD code using a Kerr
geometry and have found that, with a rapidly rotating ($a \equiv
J/J{\max} = 0.95$, $J_{\max} = GM^{2}/c$, $J$ and $M$ are angular
momentum and mass of black hole) black-hole magnetosphere. the
maximum velocity of the jet is (0.3 - 0.4) $c$.
Recently the
extraction of rotational energy from plasma in the ergosphere has been
investigated (Koide et al. 2002; Koide 2003). All of these GRMHD
simulations were made assuming axisymmetry with respect to the $z$-axis
and mirror symmetry with respect to the plane $z = 0$.  It is known
that the axisymmetric assumption suppresses azimuthal instabilities.

Magnetic fields are expected to play an essential role in the dynamics
of most relativistic objects such as active galactic nuclei, X-ray
binaries, gamma-ray bursts, and quasars. In accretion disks, the most
important role of the magnetic field is angular momentum transport,
which twists the field, and results in torsional Alfv\'{e}n waves (TAWs)
and associated jet formation (Koide et al. 1999). Based on the importance
of magnetorotational instability
(MRI) in accretion disks \citep{bah98}, the nonlinear growth of these
instabilities has been investigated by three-dimensional global MHD
simulations of black hole accretion disks
\citep{haw00,igu00,haw01,hak01,mhm00,hbs01,
arc01,raa01,hak02,hab02,kah02,mam03,igu03} using a pseudo-Newtonian
potential \citep{pacy80}. \citet{mam03} investigated black hole
accretion flow using three-dimensional global resistive MHD
simulations. They conclude that the reconnection in the innermost
region releases enough energy to explain the origin of X-ray bursts
from black hole candidates.

Recently, more GRMHD simulations have been developed by
\citet{dev02,dev03a,gam03}. In particular, \citet{gam03} have
employed the numerical scheme HARM (High Accuracy Relativistic
Magnetohydrodynamics) in which the conservation schemes for GRMHD
are used \citep{ksk98,ksk99,kmsk00}.  A nonconservative scheme for
GRMHD following a ZEUS-like approach has been developed also by
\citet{dev02}.  They also compare the code performance of 1-D and
2-D test cases.  \citet{dev03b} have investigated the properties of
accretion flows with tori in the Kerr metric using 3-D GRMHD
simulations. The MRI in torus has also been investigated using
Schwarzschild and Kerr metrics (De Villiers, Hawley, \& Krolik 2003;
Hirose, et al. 2004, De Villiers, et al. 2004; Krolik, Hawley, \&
Hirose 2004).

Different studies have chosen various initial magnetic fields. We have chosen
an initial magnetic field from the Wald's exact analytic solution for
a Kerr black hole immersed in a uniform magnetic field (Wald 1974;
Punsley 2001; Komissarov 2004).
De Villiers, Hawley \& Krolik (2003) have performed simulations
in which the initial conditions for the accretion disks and magnetic
fields are appropriate for MRI research. They have investigated the
properties of accretion flows in the Kerr metric including bi-conical
outflows in the quarter sphere with periodic boundary conditions in
the azimuthal direction. These simulations use the torus accretion
disks as an initial condition. The inner edge of the accretion disk
is located at $r = 15 r_{\rm S}$, and the inner edge moves into the
innermost stable radius within one orbit (De Villiers, Hawley,
\& Krolik 2003). In these simulations, the initial field consists of
axisymmetric (azimuthal) field loops, laid down along an isodensity
surface within the torus, which is different from the initial uniform
magnetic field used in simulations by Koide, Shibata, \&
Kudoh (1999) and Nishikawa et al. (2002, 2003). The initial weak loop
magnetic fields contribute to the generation of the MRI and
the associated jet. However, the structure of magnetic fields piled
up near the black hole in their simulations is different from the
magnetic fields which are twisted and pinched (not wound) by the free-falling
corona and accretion disk (Koide, Shibata, \& Kudoh 1998, 1999;
Koide et al. 2000; Nishikawa et al. 2002, 2003).
The absence of a free-falling corona (Bondi \& Hoyle 1944; Bondi 1952;
Michel 1972) as well as the initial loop magnetic field in
(De Villiers, Hawley \& Krolik 2003) results in different dynamics
of the accretion disk and associated jet formation, compared to the
case with the free-falling corona and initial uniform magnetic field in
Koide, Shibata, \& Kudoh (1999) and Nishikawa et al. (2002, 2003).

In this paper we present a fully three-dimensional, GRMHD simulation of
jet formation with a thin accretion disk.  Our motivation for this type
of simulation is to determine the parameters necessary for relativistic
jet formation and the resulting interaction and instabilities found
between the accretion disk and the black hole.  The simulation we
present here was performed using the same parameters as in
\citet{ksk99} in order to determine the physical differences resulting
from a full 3-D versus a 2-D simulation with axisymmetry and  mirror
symmetry at the equator. The three-dimensional simulation allows us to
study the evolution of jet formation because it is run for longer
light-crossing times than the previous two-dimensional simulations.  We
find that at the later stages of the simulation the accretion disk
becomes thick and a wind is formed with a much wider angle than the
collimated jet formed at the earlier stage.  In Section 2, we describe
the governing GRMHD equations.  The simulation model and numerical
method is described in Section 3. Our new simulation results are
presented in Section 4 followed by a summary and discussion in Section
5.

\section{3-D GRMHD Governing Equations and Numerical Approach \label{GovEqu}}

The governing equations for GRMHD are written in conservation form
based on \citet{wei72} and \citet{mtw73}.
Four-dimensional space-time
parameters are written with Greek indices, while Latin indices are used
for three-dimensional space, $x^\alpha = (ct,x^j)$.  The physical
parameters include, four-velocity ($U^\alpha$), proper density
($\rho$), proper pressure ($p$), adiabatic index ($\Gamma$), and proper
internal energy density ($e_{\rm int}\equiv \rho c^{2} +p/(\Gamma
-1)$).  Relativistic
parameters include the Christoffel symbols ($\Gamma^\alpha_{\beta
\gamma}$) and metric terms ($g_{\alpha \beta}$).  The pressure is
related to the other variables by the equation of state, $p \propto
\rho^{\Gamma}$.  The specific relativistic enthalpy is given by, $h =
e_{\rm int} +p = \rho c^{2} +\Gamma p/(\Gamma -1)$.

The conservation equations are derived from the continuity equation,
$\nabla _\gamma  \left( \rho U^\gamma \right) = 0$, and the
conservation of the stress energy tensor, $\nabla _\mu  T^{\mu \upsilon
}  = T^{\mu \upsilon } _{;\mu }  = T^{\mu \upsilon } _{,\mu } +
T^{\alpha \upsilon } \Gamma ^\mu  _{\alpha \mu }  + T^{\mu \alpha }
\Gamma ^\upsilon  _{\alpha \mu }  = 0$.  From Maxwell's equations, the
electromagnetic field-strength tensor is denoted by $F_{ \alpha \beta}
= \partial_\alpha A_\beta - \partial_\beta A_\alpha$ and the
four-current density by $J^\beta$, where $A_\alpha$ is the four-vector
potential.  The Maxwell equations are written $\partial_\mu F_{ \alpha
\beta} + \partial_\alpha F_{ \beta \mu} + \partial_\beta F_{ \mu
\alpha} = 0$ and $\nabla _\alpha  F^{\alpha \beta} = -J^\beta$.

For the simulation in the context of this paper, we have assumed
infinite electric conductivity, $F_{\alpha \beta} U^\alpha = 0$.  The
stress energy tensor ($T^{\alpha \beta }$) for an ideal fluid in an
electromagnetic field and in a general relativistic environment is,
\begin{equation}
T^{\alpha \beta }  = \left( \rho + p \frac{\Gamma}{\Gamma - 1}\right)
U^\alpha  U^\beta
+ pg^{\alpha \beta }
+ F^\alpha_\sigma F^{\beta \sigma} - g^{\alpha \beta} \frac{F^{\lambda \gamma}
F_{\lambda \gamma}}{4}.
\end{equation}
We assume that the off-diagonal spatial elements of the metrics $g_{\mu
\nu}$ vanish: $g_{ij} =0 \, (i\ne j)$.
Here Roman indeces ($i, j$) run from 1 to 3, the
\begin{equation}
     g_{00} = -h_{0}^{2},   \; \; \; g_{ii} = h_{i}^{2}, \; \; \;
     g_{i0} = g_{0i} = -h_{i}^2\omega_{i}/c,
\end{equation}
The line element is written
\begin{equation}
   ds^{2} = -\alpha^{2}(cdt)^{2} + \sum_{i=1}^3 (h_{i}dx^i -c\beta^i\alpha dt)^{2}.
\end{equation}
Here, the lapse function and shift velocity are defined by $\alpha =
\sqrt{h^{2}_0 + \sum_{i=1}^3 (\frac{h_i\omega_i}{c})^{2}}$ and $\beta^i =
\frac{h_i\omega_i}{c\alpha}$, respectively.

The 3 $+$ 1 form of equations are employed (Weinberg 1972; Thorne,
Price, \& Macdonald 1986). The 3 $+$ 1 set of equations used in these
simulations for the general relativistic conservation laws governing the
plasma and Maxwell equations were derived in Koide, Shibata, \& Kudoh
(1999) and Koide (2003). The detailed presentation of the equation is
provided in Koide (2003) and will not be repeated here. The numerical
scheme is constructed in the set of conservation variables  $D {\rm
(density)}, P^i {\rm (momentum)}, \epsilon$ (energy), and $B^i$ (for
details see Koide, Shibata, \& Kudoh (1999) and Koide (2003)).  The
values of the primitive variables ${\bf v}$, $\rho$, Lorentz factor,
and $p$ are calculated from  $D, P^i, \epsilon$, and $B^i$. Two
nonlinear, simultaneous algebraic equations with unknown variables, $a
\equiv \gamma -1$ and $b \equiv \gamma({\bf v}\cdot {\bf B})/c^{2}$, are
solved using a two variable Newton-Raphson iteration method (for
details see Koide, Shibata, \& Kudoh (1999) and Koide (2003)). Here the
Lorentz factor is defined $\gamma = 1/\root \of{(1 - \sum_{i=1}^3
(v^i/c)^{2})}$.

In the course of the simulation, Gauss's law for the magnetic field,
$\nabla\cdot{\bf B} = 0$, and the charge conservation law,
$\nabla_{\mu}{\bf J}^{\mu} =0$ must be verified. Here, ${\bf J}^{\mu} =
(c\rho_{\rm c}, J^{1}, J^{2}, J^{3})$ is the current four vector. We adapt
the Boyer-Lindquist coordinate system ($t, r, \theta, \phi$), the
metric of Schwarzschild space-time is simplified to
\begin{equation}
     h_{0} = \alpha,   \; \; \; h_{1} = 1/\alpha, \; \; \; h_{2} = r, \; \; \;
     h_{3} = r\sin\theta, \; \; \; \omega_{i} = 0
\end{equation}
where the lapse function is $\alpha = (1 - r_{\rm S}/r)^{1/2}$ and
Schwarzschild radius is $r_{\rm S} = 2GM/c^{2}$ (${\rm G} =
6.67\times10^{-11}$ Nm$^2/$kg$^2$ is the gravitational constant). Here,
$r,\, \theta,\, \phi$ are radial, latitudinal, and azimuthal
coordinates, respectively.  The line element for the Schwarzschild
metric has the form
\begin{eqnarray}
ds^{2} = -\left( 1- \frac{r_{\rm S}}{r} \right)c^{2} dt^{2} +
  \left(1- \frac{r_{\rm S}}{r} \right)^{-1} dr^{2} + r^{2} d\theta^{2}
   + r^{2} \sin^{2} \theta d\phi^{2}.
\end{eqnarray}
We use the modified {\it tortoise} radial coordinate, $r' = {\rm
ln}(r/r_{\rm S} -1)$.

The GRMHD equations described in this section, with the applied initial
and boundary conditions (which will be described in the next section),
were solved using the simplified total variation diminishing (TVD)
method. The simplified TVD method was developed by Davis (1984) for
studying violent phenomena such as shocks (Koide, Nishikawa, \& Mutel
1996; Nishikawa et al. 1997, 1998; Koide, Shibata, \& Kudoh 1999). This
method stems from Lax-Wendoroff's method, with additional diffusion
terms.  It was verified in \citet{ksk99} that this method obeys the
laws of conservation of energy. The conservation laws ($\partial
\rho_{e}/ \partial t + \nabla \cdot \alpha (J+\rho_{e} c \beta) = 0$),
where $\alpha$ and $\beta$ is lapse function
and shift vector, respectively. is restraint almost zero within the
round-off error 
with the double precision.  It was also confrmed that when we take
the nonrelativistic limit ($(v/c)^{2} \ll 1$), the Newtonian
calculation (nonrelativistic MHD simulation) of a magnetized jet
resulted (Koide, Nishikawa, \& Mutel 1996).

\section{3-D GRMHD Simulations: Simulation Parameters and Numerical Techniques
   \label{conditions}}

In this section we present our results showing the fully three
dimensional simulations of jet formation using general relativistic
magnetohydrodynamics. In the assumed initial state of our simulations,
the region is divided into two parts: a background corona surrounding
the black hole, and an accretion disk along the equatorial plane of the
geometry (see Fig. 1a).  The simulation results are displayed in the
Cartesian coordinate transforming from the Boyer-Lindquist coordinates
($t,\, r,\, \theta, \phi$) coordinate ($x = r\sin \theta \cos \phi, y =
r\sin \theta \sin \phi, z = r \cos \theta$).  The black hole is located
at the origin ($x = y = z = 0$).  The coronal plasma is set in a state
of transonic free-fall flow (Bondi accretion), with an adiabatic
exponent, $\Gamma = 5/3$, specific relativistic enthalpy, $h=1.3$, and
sonic point located at $r=1.6 r_{\rm S}$.  The Keplerian disk region is
located at $ r > r_{\rm D} \equiv 3r_{\rm S}, | {\rm cos} \theta | <
\delta, $ where $\delta = 1/8$, $r_{\rm D}$ is the disk radius and
$r_{\rm S}$ the Schwarzschild radius.  In this region the density is
100 times that of the background corona, while the orbital velocity,
$v_\phi$, is relativistic and purely azimuthal:  $v_\phi = v_{\rm K}
\equiv c/[2(r/r_{\rm S} -1)]^{1/2}$, where $v_{\rm K}$ is the Keplerian
velocity.  (Note that this equation reduces to the Newtonian value,
$v_{\rm \phi} = \sqrt{GM/r}$, in the non-relativistic limit $r_{\rm
S}/r \ll 1$).  The pressure of both the corona and the disk are assumed
to be equal to that of the transonic solution.  The initial conditions
for the plasma surrounding the black hole in terms of the transonic
flow velocity, density and pressure ($v_{\rm ffc}, \rho _{\rm ffc}, p
_{\rm ffc}$) and the disk properties ($v_{\rm dis}, \rho _{\rm dis}, p
_{\rm dis}$) are:
\begin{equation}
\rho = \rho _{\rm ffc} +\rho _{\rm dis}
\end{equation}
\begin{equation}
\rho _{\rm dis} = \left \{  \begin{array}{cc}
100 \rho _{\rm ffc} &
\verb!   ! ( r > r_{\rm D} \verb!  ! {\rm and} \verb!  !
|{\rm cot} \theta| < \delta )\\ 0                  &
\verb!   ! ( r \leq r_{\rm D} \verb!  ! {\rm or} \verb!  !
|{\rm cot} \theta| \geq \delta )
\end{array} \right .
\end{equation}
\begin{equation}
(v_r, v_\theta , v_\phi) = \left \{  \begin{array}{cc}
(0, 0, v_{\rm K}) &
\verb!   ! ( r > r_{\rm D} \verb!  ! {\rm and} \verb!  !
|{\rm cot} \theta| < \delta )\\ (-v_{\rm ffc}, 0, 0)        &
\verb!   ! ( r \leq r_{\rm D} \verb!  ! {\rm or} \verb!  !
|{\rm cot} \theta| \geq \delta )
\end{array} \right .
\end{equation}
where we set $\delta = 0.125$ and the smoothing length is $0.3 r_{\rm
S}$. The initial conditions for the free-falling corona are based on
theories by Bondi \& Hoyle (1944), Bondi (1952), and Michel (1972).

An initially uniform magnetic field is applied perpendicular to the
accretion disk.  It is set to the Wald solution (Wald 1974), which
represents the uniform magnetic field around a Schwarzschild black
hole; $B_r = B_0 {\rm cos} \theta$ and $B_{\theta } = - \alpha B_0 {\rm
sin} \theta $ (where $\alpha$ is the lapse function defined in section
2).  At the inner edge of the accretion disk, the proper Alfv\'{e}n
velocity is $v_{\rm A} = 0.015c$ when $B_0 =0.3 \, \root \of {\rho
_0c^{2}}$.  The Alfv\'{e}n velocity in the fiducial observer frame is
\begin{equation}
v_{\rm A} \equiv \frac{B}{\root \of {\rho + [\Gamma p/(\Gamma -1)+B^{2}]/c^{2}}}.
\end{equation}
The plasma beta of the corona, which is defined as the ratio of gas
pressure to magnetic pressure, is: $\beta \equiv p/(B^{2}/2)=1.40$ at
$r=3r_{\rm S}$. The sound speed is calculated from $v_{\rm S} \equiv
c(\Gamma p /h)^{1/2}$.  Our simulation domain is, $1.1 r_{\rm S} \leq r
\leq 20 r_{\rm S}$, $0 \leq \theta \leq \pi$, $0 \leq \phi \leq 2\pi$,
with a mesh size of $100 \times 60 \times 120$.  The effective linear
mesh spacing at $r=1.1r_{\rm S}$ and at $r=20r_{\rm S}$ is $5.38 \times
10^{-3} r_{\rm S}$ and $0.97 r_{\rm S}$, respectively.  The angular
spacing along the polar and azimuthal directions is $5.2 \times
10^{-2}$ rad. At the $z$-axis ($\theta = 0, \; \pi$) the fixed boundary
is used, therefore the kink instability is suppressed near the axis.
(However, a warping instability near the equator is not suppressed.)
The following radiative boundary condition is imposed at $r=1.1 r_{\rm
S}$ and at $r=20 r_{\rm S}$:
\begin{equation}
u_0^{n+1} = u_0^n + u_1^{n+1} - u_1^n ,
\end{equation}
where the superscript $n$ denotes the time step, the subscript 0 refers
to the boundary node, and the subscript 1 is the node neighboring the
boundary.  The computations were done on an SGI ORIGIN 2000 system with
0.898 GB of internal memory.  47 hours of CPU time were needed for
10,000 time steps.

Contributions to jet formation from the gas pressure and
electromagnetic field components are found using the following
equations derived in \citet{ksk99}:
\begin{equation}
W_{gp} \equiv -{\bf v} \cdot \nabla p
\label{PowerGP}
\end{equation}
\begin{equation}
W_{EM} \equiv {\bf v} \cdot \left({\bf E} + {\bf J} \times {\bf B} \right)
\label{PowerEM}
\end{equation}
These equations will be used for showing the energy sources for
generating the jet in Figs. 6b and 7b.

\section{Simulation Results: Jet Formation}

The evolution of jet formation for the simulation performed in the
region previously described ($1.1 r_{\rm S} \le r \le 20 r_{\rm S}$, $0
\le \theta \le \pi$, and $0 \le \phi \le 2\pi$ with a grid resolution
of $100 \times 60 \times 120$) is shown in Figure 1.  The physical
values calculated in spherical coordinates are transformed to Cartesian
coordinates to display their values (for only limited region near the
black hole ($|x|, |y|, |z| \le 8r_{\rm S}$).  The black hole is
located at the origin ($x = y = z = 0$).  The parameters used in this
simulation are the same as those of the axisymmetric simulations shown
in Fig. 6 of Koide, Shibata, \& Kudoh (1999). Figure 1 shows the
evolution of the proper density on a logarithmic scale (colored
shading) with the flow velocity (shown by the directional arrows) in
the $x - z$ plane ($y = 0$).  The maximum values are normalized for all
panels. The black circle represents the black hole's Schwarzschild
radius. Figure 1a presents the initial conditions described in Section
\ref{conditions}.  Figure 1b ($t= 39.2 \tau_{\rm S}$) shows that a
shock forms in the region of $x = 3.5r_{\rm S}$. This shock is due to
the rapidly infalling material reaching the centrifugal barrier located
at $x = 2r_{\rm S}$.  At this time, the initial signs of jet formation
are seen.  When the simulation reaches $t= 60.0 \tau_{\rm S}$
(Fig.\ 1c), the jet structure is clearly seen. The formed jet is
identified by higher density (Fig. 1c) and pressure (Fig. 2c) along the
vertical line at $x = \pm 4r_{\rm S}$. The arrows show the velocity of
the jet. This phenomenon was also seen in previous three-dimensional
simulations (Nishikawa at al.\ 2002, 2003). In the last frame (Fig. 1d)
at $t = 128.9\tau_{\rm S}$, the accretion disk has thickened and the
jet appears to fade into a weak wind that flows near the accretion
disk. Symmetry at the equator is seen as expected even with the freedom
of the three-dimensional simulation as no asymmetric perturbation has
been applied.

The proper pressure, over the same time scale, is shown in Figure 2.
The jet forms in the shape of a hollow cylinder which is shown clearly
in this figure.  The pressure in this region is high due to the shock
and adiabatic heating at the boundary of the decelerated flow. This
high pressure contributes greatly to the jet formation (see Equation
\ref{PowerGP}), which is shown by the green curves in Figs. 6b and 7b.
A bi-conical high pressure region with velocity arrows directed away
from the disk in Figure 2c represents a jet.
However, at the later time the jet is quenched and the pressure
decreases, as the shock disappears at $r=3r_{\rm S}$. Figure 2d shows
the faded jet and the wind generated along the disk.

The evolution of the ratio of gas to magnetic pressure $\beta =
p/(0.5B^{2})$ in the $x - z$ plane ($y = 0$) is shown in Fig. 3. In
this figure the maximum and minimum values are not normalized in
order to see the relative values at each time. The initial cold
accretion disk is heated and the shock is created around $x/r_{\rm
S} = \pm 3$ as shown in Fig. 3b. The generated conical jet is seen
in Fig. 3c. This structure is also found in tori simulations (Fig. 2
in Hirose et al.\ (2004)). The jet is ejected along the twisted
magnetic field (shown in Fig. 4c). Two layers of high $\beta$
regions are found near the black hole. After the jet is faded, the
accretion disk is broadened as seen in Fig. 3d.

The evolution of jet formation is closely related to the deformed
magnetic field as shown in Fig. 4. The magnetic field lines, which are
initially perpendicular to the accretion disk (Fig 4a), are twisted by
the accretion disk as the simulation progresses and are pinched by the
falling corona and accretion material (Fig. 4b). Later the disk drags
the magnetic field further in the azimuthal direction, transferring
angular momentum outward even as matter falls toward the black hole.
The pinching twisting field lines increase the magnetic field pressure
and generate a jet near the black hole (Fig. 4c). Equation
\ref{PowerEM} presents the power of the electromagnetic forces which
drives jet formation as is also shown by the blue curves in Figs. 6b
and 7b. At $t = 128.9 \tau_{\rm S}$ the magnetic field has relaxed and
is consistent with what is seen in Figures 1d and 2d where the jet
fades.

Recently, Hirose et al.\ (2004) have analyzed the magnetic field
structures found in 3-D GRMHD simulations of tori in the Kerr metric
(De Villiers, Hawley, \& Krolik 2003). It should be noted that their
different initial magnetic field conditions lead to different
magnetic field dynamics. For example, the sample field lines in Fig.
6 in Hirose et al.\ (2004) for the spin parameter $a/M = 0.9$ (even
$a/M = 0$) are substantially different from the magnetic field lines
in Fig. 4. Since the magnetized disk in our model is stable to MRI,
the magnetic fields
are never tangled. De Villiers et al. (2004) have found the unbound  
outflows in the axial funnel region, which consist of two
components: a hot, fast, tenuous outflow in the axial funnel proper,
and a colder, slower, denser jet along the funnel wall.

The time required to form the jet in this three-dimensional simulation
is similar to the two-dimensional axisymmetric simulations.
Ultimately, we expect that the dynamics of jet formation are modified
by the additional coordinate freedom in the azimuthal dimension (when
axisymmetry with respect to the $z$-axis is removed), such as the
difference between axisymmetric 2-D and full 3-D MHD simulations found
by Matsumoto (1999). Due to the growth of non-axisymmetric
instabilities, the magnetic field lines were compressed into helical
bundles in the jet. In this simulation with a thin disk a global and
visible instability was not found in the $x-y$ plane. One of the
possible reasons is that the strong magnetic field along the $z$-axis
(low $\beta$ value) prevents growth of MRI. The magnetic fields found
with MRI are usually along the azimuthal direction (the initial $\beta$
value is larger) (e.g. De Villiers, Hawley, \& Krolik 2003).

In order to analyze the properties of jet formation in this
simulation, we present the physical variables using several
one-dimensional slices along the radial axis and the jet formation
axis. Figures 1c and 2c show the density and pressure at time
$60.0\tau _{\rm S}$ corresponding to the time of our one-dimensional
slices shown in Figs. 5 - 7. At this time the generated jet has
maximum velocity. The parameters for these slices where chosen in
order to perform comparisons with \citet{ksk99}, and are plotted
showing one dimensional radial ($x$-axis) slices along $z = 0$, $z =
5.6r_{\rm S}$ and $x = 4.48r_{\rm S}$ in the Cartesian coordinate
(for all figures at $y = 0$).

The parameters along the $x$-axis (at $y = z = 0$) with the event
horizon at $x/r_{\rm S} = \pm 1.0$ are shown in Figure 5. The
central gaps in these panels correspond to the black hole region.
Figure 5a which shows the density (blue), pressure (green) and
$z$-component magnetic field energy $B^{2}_{\rm z}/2$ (red) reveals
a shock around $x =\pm  3.0 r_{\rm S}$ outside the centrifugal
barrier at $x =\pm  2.0 r_{\rm S}$. Shock and adiabatic heating due
to strong deceleration of the flow yields a relativistic region
within $x = \pm 2.0 r_{\rm S}$ where $p \approx \rho c^{2}$.  Figure
5b shows the power contributions from the gas pressure (green) and
electromagnetic field (blue) components independently as depicted in
Equations \ref{PowerGP} \& \ref{PowerEM}.  These energies contribute
to decelerating the accretion disk plasma (attention must be given
to the sign ($-W$)). The electromagnetic force is produced by the
magnetic tension caused by the strongly pinched magnetic field lines
as shown in Fig. 4c. The individual magnetic field components are
shown in Fig. 5c. The $z$-component $B_{\rm z}$ (red) is pinched by
the falling corona and accretion disk. However, other components
($B_{\rm x}$ (blue) and $B_{\rm y}$ (green)) are close to zero.
Slight humps in $B_{\rm x}$ (blue) at $x =\pm 2r_{\rm S}$ are found
near the black hole but further investigation is necessary to
ascertain the cause.  The three-velocity components $v_{\rm x}$
(blue), $v_{\rm y}$ (green), and $v_{\rm z}$ (red) are plotted in
Fig. 5d. The value of $v_{\rm y}$ reverses at the location of the
centrifugal barrier at $x = \pm 2r_{\rm S}$. The inner edge of the
disk enters into the unstable orbit region, $|x| \le 3r_{\rm S}$, as
shown in Fig. 5d.  The material accretes to the points around $x =
\pm 2r_{\rm S}$ but is slowed down by the compressed magnetic field
($B_{\rm z}$ (red)) as seen in Fig. 5c.  The advection flow in the
unstable orbit region stops at $x \approx \pm 2r_{\rm S}$, and the
plasma is ejected in the $z$-direction as shown in Figs. 1c and 2c.
Inside the barrier, the plasma is accelerated rapidly to a
relativistic velocity near the horizon as seen in Fig. 5d ($v_{\rm
x}$ (blue)). This highly heated, high density region should
contributes to radiation which can be observed by present and
planned space based observatories (see Fabian et al.\ 2000; Reynolds
and Nowak 2003).

Figure 6 shows the same parameters as in Figure 5, but along the $z =
5.6r_{\rm S}$ slice ($y = 0$).  Since this slice is well above the
Schwarzschild radius of the black hole, there is not a central gap in
the figures.  Figure 6b clearly shows the separate layers of the jet
resulting from the gas pressure (green around $x = \pm 4r_{\rm S}$) and
the electromagnetic forces (blue around $x = \pm 5r_{\rm S}$). A
double-layered jet is found where the inner part of the jet is a {\it
gas-pressure-driven jet} and the outer one is a {\it magnetically
driven jet} (Koide, Shibata, \& Kudoh 1999). From the velocity profiles
in Figure 6d, the jet appears to have a width from $3 r_{\rm S} \le |x|
\le 5 r_{\rm S}$. The electromagnetically driven component has the
larger magnetic field $B_{\rm y}, B_{\rm z}$ and a large azimuthal
velocity $v_{\rm y}$. It should be noted that at this time ($\tau_{\rm
S} = 60.0$) a torsional Alfv\'en wave is excited (Koide, Shibata,
\& Kudoh 1999). Later the outwards
propagation of the TAW contributes to formation of a wind, and as
will be discussed later, the peak of $v_{\rm z}$ will move outward.

Figure 7 presents the physical variables along the $x = 4.48r_{\rm
S}$ slice ($y = 0$). This slice is along the vertical axis of the
jet rather than the horizontal plane of the disk as in the previous
figures (Figs. 5 and 6). The density (blue) and pressure (green)
increases are expected around the accretion disk in Fig.  7a. The
equatorial plane is located at $z = 0$ and the density in the disk
is much higher than other values as shown in Fig. 7a.  Figure 7b
shows the contribution to jet generation due to the electromagnetic
field ($2 < |z/r_{\rm S}| < 4$) (outer part) and gas pressure ($5 <
|z/r_{\rm S}| < 8$) (inner part), and is consistent with the
increase of the jet $v_{\rm z}$ (red) in Fig. 7c. However, the
region of acceleration due to gas pressure is smaller than in the
axisymmetric case. This may come from the fact that the time here is
somewhat later than the $t = 52\tau_{\rm S}$ in the axisymmetric
case and/or due to 3-D effects.  The three-velocity components are
plotted in Fig. 7c. It should be noted that the jet has a larger
azimuthal component ($v_{\rm y}$) near the disk than the $z$
component ($v_{\rm z}$) which becomes larger far from the disk. A
slightly positive radial velocity $v_{\rm x}$ (blue) is found far
from the disk in Fig. 7c, and this indicates outward angular
momentum transfer. The Alfv\'{e}n velocity $v_{\rm A}$ and the sound
velocity $v_{\rm s}$ are shown along with the absolute value of
$v_{\rm z}$ in Fig. 7d. The velocity $v_{\rm z}$ gradually increases
in the magnetically driven region $5r_{\rm S} \le z \le 7r_{\rm S}$
but does not exceed either the sound or Alfv\'{e}n velocities.

Figure 8 shows one dimensional slices of the velocities $v_{\rm x}$
(blue), $v_{\rm y}$ (green), and $v_{\rm z}$ (red), at $z = 5.6r_{\rm
S}$ ($y = 0$) for times (a) $t=0.0\tau _{\rm S}$, (b) $t=39.2\tau _{\rm
S}$, (c) $t=60.0\tau _{\rm S}$, and (d) $t=128.9\tau _{\rm S}$. The jet
is seen in $v_{\rm z}$ (red curve) and its maximum velocity can be
found in Fig. 8c. Comparison between the velocity structure at $t =$
$60\tau_{\rm S}$ (Fig. 8c) and $80\tau_{\rm S}$ (not shown) reveals
that the jet velocity  ($v_{\rm z}$) and the azimuthal velocity
($v_{\rm y}$) are smaller at $t = 80\tau_{\rm S}$. Additionally, the
location of peak velocities ($v_{\rm z}$) are shifted outward at the
later time.  This may come from the fact that the wind is caused by
the unwinding magnetic fields with the TAWs which are propagating
outward in the disk. These
results are consistent with the fact that the jet has faded and a weak
wind is generated around $x = 7 r_{\rm S}$ as seen in Fig. 8d.
Furthermore, the rotation of the accretion disk ($v_{\rm y}$) becomes
very small and rotates slightly in the reverse direction around $x =
\pm 3.5r_{\rm S},$ and $\pm 7r_{\rm S}$. We plan to investigate how
these changes are affected by inflow of matter to the system in the
future.

A comparison of our present results with previous two-dimensional
results shown in \citet{ksk99} reveals some minor differences that
may be due to the three-dimensional modeling.  The most prominent
difference between this 3-D simulation and previous 2-D simulations
lies in the longer run time and hence shows a significant change in
the disk-jet system to a disk-wind system (Fender 2002, 2003). This
transition may be caused by outward TAW propagation in the disk. The
jet is mainly produced by the gradient of gas pressure enhanced by
the shock around $r =3r_{\rm S}$, as shown  in Figs.  1c, 3c and 6b.
After $t=60\tau_{\rm S}$, the TAW moves outward in the disk and is
weakened.  The wind appears while the TAW is moving outward.  After
outward propagation of the TAW, the pressure of the outer disk
becomes larger than the initial condition (Fig. 2a). Therefore, the
wind may be caused by unwinding of the twisted and pinched magnetic
fields due to the decreased accreting matter, which creates the TAW
propagating outward in the disk. This transient change may be
controlled by the mass accretion rate onto the black hole. It should
be noted that in the simulations with initial weak azimuthal
magnetic fields the magnetic fields are tangled as shown in Fig. 6
in Hirose et al.\ (2004) as a result of MRI where TAWs are not
excited.

\section{Summary and Discussion}

We have presented a fully 3-D GRMHD simulation of a Schwarzschild
black hole accretion disk system with moderate resolution.  Because
no deliberate asymmetric perturbation was introduced, the present 3D
simulation remains largely axisymmetric. While we found some
numerical differences between this simulation and previous
two-dimensional simulations, our results reaffirm the previous
result of a two-layered jet. Our results suggest that the jet
develops as in the 2-D case. Our longer simulation has allowed us to
study jet evolution.  At the end of this simulation the jet fades
and a weak wind is generated by a thickened accretion disk. This
phenomena was not observed in the two-dimensional simulations
because they did not run for this duration.

In this simulation at the outer boundary of the accretion disk no
matter is injected, therefore after accreted matter is ejected from
the accretion as a jet, the power to generate a jet is dissipated
and the jet is switched to a wind. However, if streaming matter is
injected at the outer boundary of accretion disk, transient changes
may be controlled by accretion rates with instabilities and be
related to the state transitions. Such disk-jet coupling in black
hole binaries is reviewed by Fender (2002, 2003) and Fender,
Belloni, \& Gallo (2004). Black hole binaries exhibit several
different kinds of X-ray `state'. The two most diametrically
opposed, which illustrate the relation of jet formation to
accretion, can be classified as low/hard/off and high/soft states
(Fender 2003). These two states provide different luminosity.
Pellegrini et al.\ (2003) have discussed a nuclear bolometric
luminosity and an accretion luminosity $L_{\rm acc}$ in terms of the
accretion rate $\dot{M}$ and jet power. Clearly these issues can be
investigated by further 3-D GRMHD simulations, and future
simulations will investigate jet formation with different states of
the black hole including streaming material from an accompanying
star (e.g., Miller et al.\ 2003).

Additionally, we plan to study the development of instabilities along
the azimuthal direction such as the proposed accretion-ejection
instability (AEI) (e.g. Tagger \& Pellat 1999; Caunt \& Tagger 2001;
Rodriguez et al. 2002; Vani\'{e}re, Rodriguez \& Tagger 2002),  the
magnetorotational instability (MRI) (e.g. Balbus \& Hawley 1991,
1998), and the ``screw'' (current-driven helical kink) instability (e.g.
Li 2000).  Such simulation studies may indicate how the observed
variabilities in AGN jets (Blandford 2003; Mirabel 2003) and
microquasars (e.g. Eikenberry et al. 1998; Ueda et al. 2002; Nobil
2003; Rau, Greiner, \& McCollough 2003) can be produced.
Initial magnetic field configurations affect the dynamics of
accretion disks and the associated phenomena such as jet formation and
its variabilities, and thus we plan to study different initial field
configurations.

The Fe K$\alpha$ fluorescent emission line in active galactic nuclei
(AGN) is believed to originate in a relativistic accretion disk around
a black hole (e.g. see reviews Fabian et al. 2000; Reynolds \& Nowak
2003). The X-ray continuum emission of MCG-6-30-15 is highly variable
(e.g. Vaughan \& Fabian 2003). The source spectrum and variablity can
be explained by two-components consisting of a variable power law
component (PLC) and an almost constant reflection-dominated component
(RDC) containing  the iron line (Fabian \& Vaughan 2003). The region
where the flare (or multiple flares) which created the PLC must be
located close to the rotation axis and within 3-4 gravitational radii
to reproduce the highly variable behavior, (Miniutti et al. 2003). Since
this behavior is most likely powered by magnetic field from the disk,
possibly linking to the hole (e.g. Blandford \& Znajek 1977; Wilms et
al. 2001), and perhaps forming the base of jet, it is essential to
perform GRMHD simulations. These issues are being investigated with
3-D GRMHD simulations with a Kerr black hole with streaming matter
from an accompanying star.

\acknowledgments
K.N is a NRC Senior Research Fellow. K.N thanks M. Aloy and D. Meier for fruitful
discussion.
K.N. is partially supported by NSF ATM 9730230,
ATM-9870072, ATM-0100997, and INT-9981508. The simulations
have been performed on ORIGIN 2000 and IBM p690 at NCSA which is
supported by NSF.

\clearpage


\begin{figure}[ht]
\epsscale{1.00}
\vspace*{-5.0cm}
\plotone{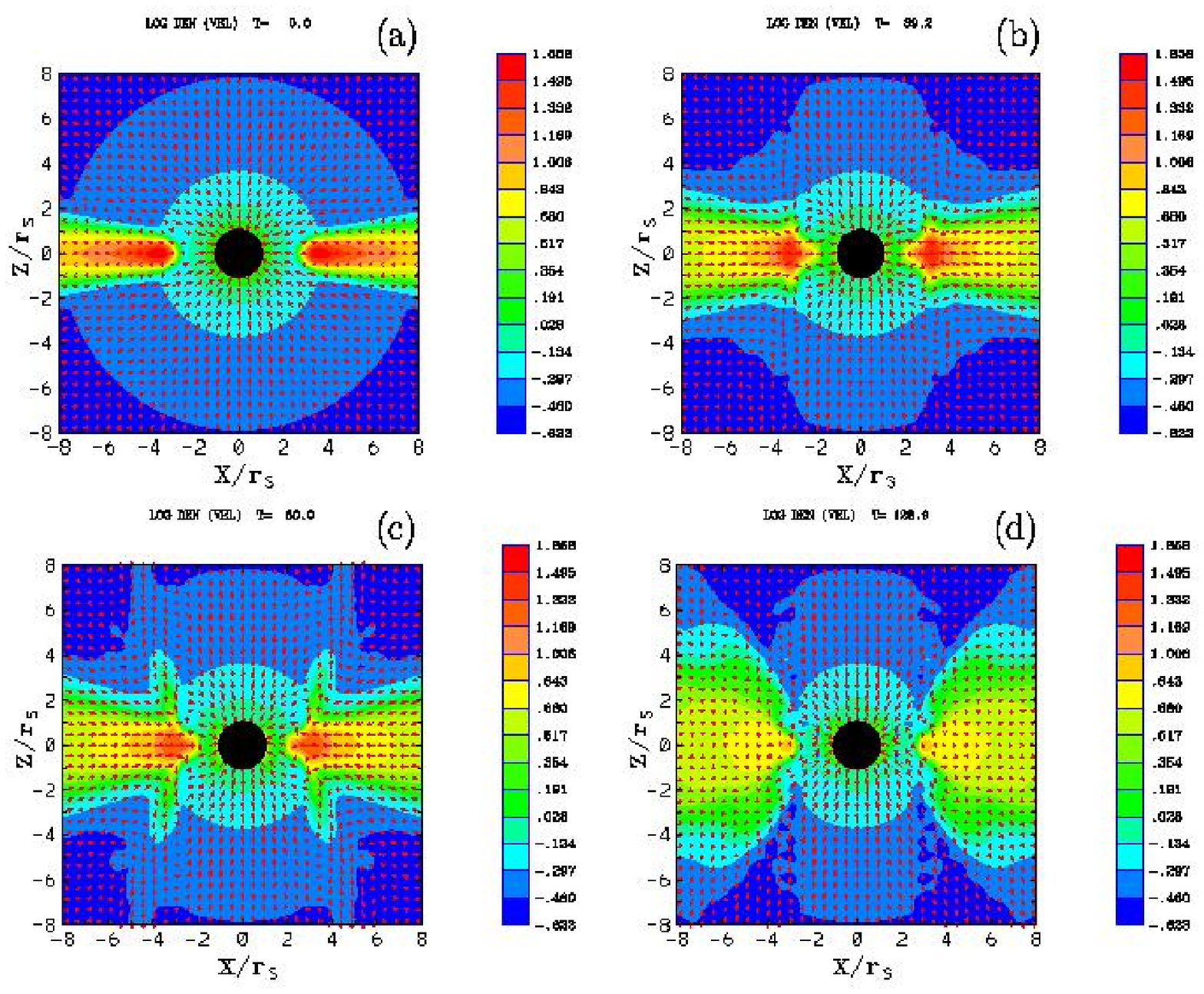}
\vspace*{-6.0cm}
\caption{The panels present the time
evolution of the proper mass density (color) with velocity arrows ($v_{x},
v_{z}$) in a transonic free-fall (steady-state falling) corona with an
initially uniform magnetic field, at (a) $t=0.0\tau _{\rm S}$,
(b) $t=39.2\tau _{\rm S}$, (c) $t=60.0\tau _{\rm S}$, and (d)
$t=128.9\tau _{\rm S}$. 
The black hole's
Schwarzschild radius is presented by the central black circle in each panel.
The jet is fully formed at $t=60.0\tau _{\rm S}$ (c). At the later time
the wind is formed with a wider angle.}
\label{densityfig}
\end{figure}

\begin{figure}[ht]
\epsscale{1.00}
\vspace*{-5.0cm}
\plotone{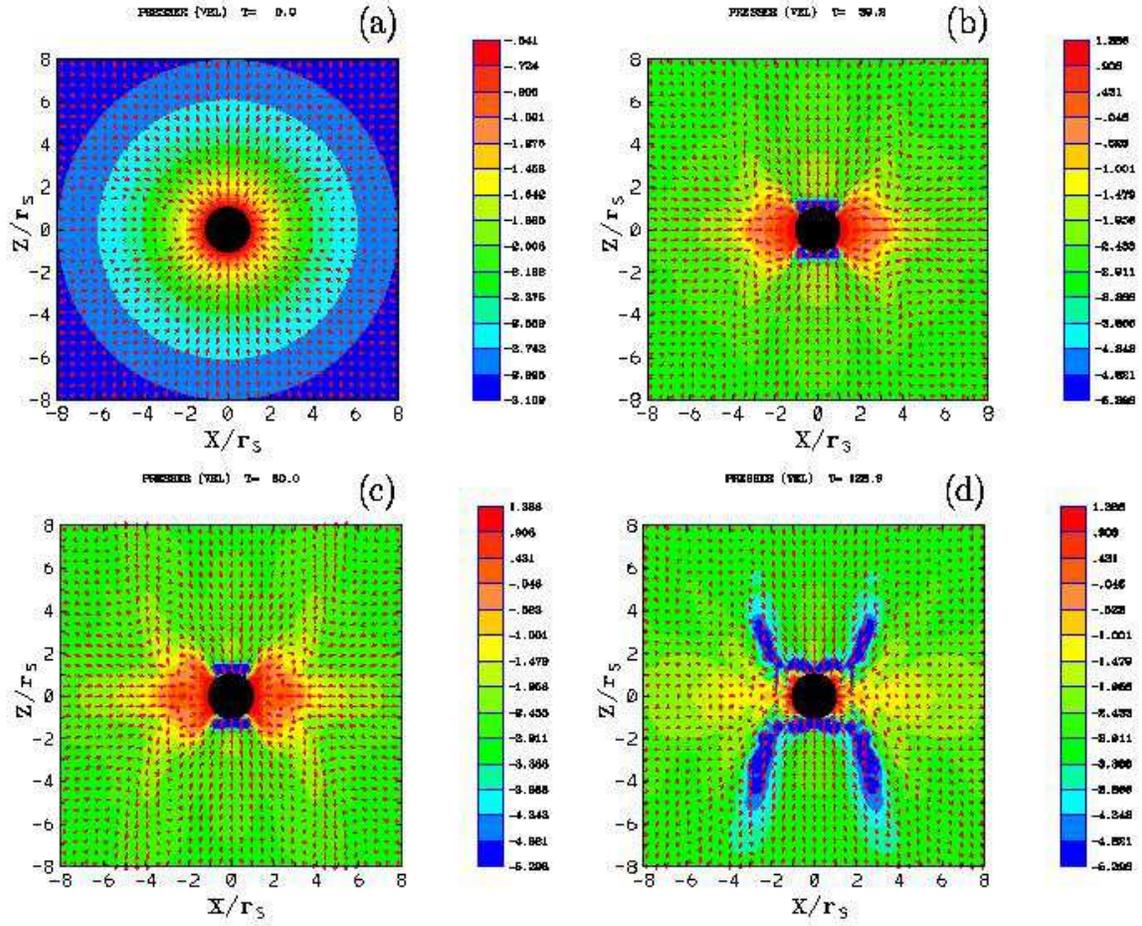}
\vspace*{-6.0cm}
\caption{These panels present the time
evolution of the proper pressure (color) with velocity arrows ($v_{x}, v_{z}$)
in a transonic free-fall (steady-state falling) corona with an
initially uniform magnetic field, at (a) $t=0.0\tau _{\rm S}$,
(b) $t=39.2\tau _{\rm S}$, (c) $t=60.0\tau _{\rm S}$, and (d)
$t=128.9\tau _{\rm S}$.}
\label{pressurefig}
\end{figure}

\begin{figure}[ht]
\vspace*{-6.5cm}
\plotone{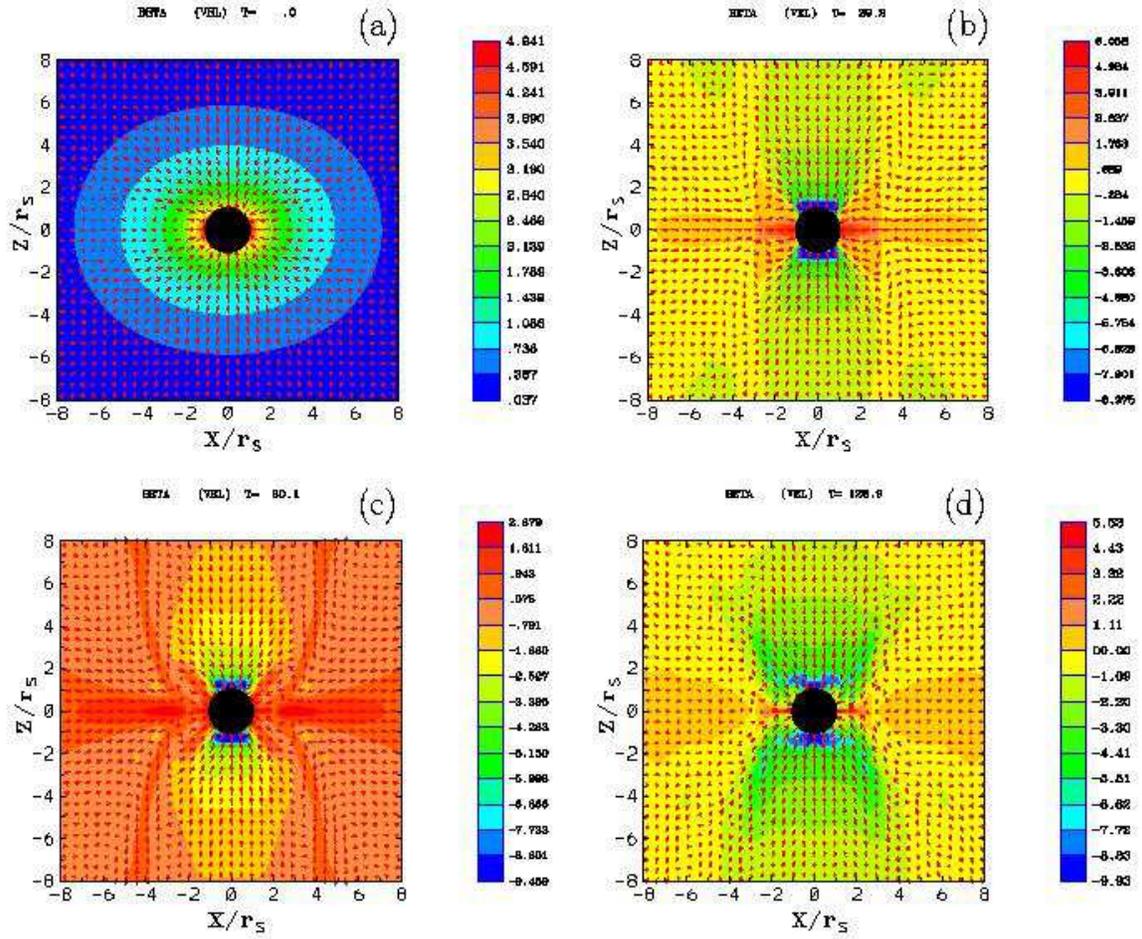}
\vspace*{-6.0cm}
\caption{The time evolution of gas to magnetic pressure $\beta = p/(0.5B^{2})$
(color) with velocity arrows at (a) $t=0.0\tau _{\rm S}$,
(b) $t=39.2\tau _{\rm S}$, (c) $t=60.0\tau _{\rm S}$, and (d)
$t=128.9\tau _{\rm S}$. The profiles of generated jets are seen clearly in
Fig. 3c.}
\end{figure}

\begin{figure}[ht]
\epsscale{1.00}
\vspace*{-4.0cm}
\plotone{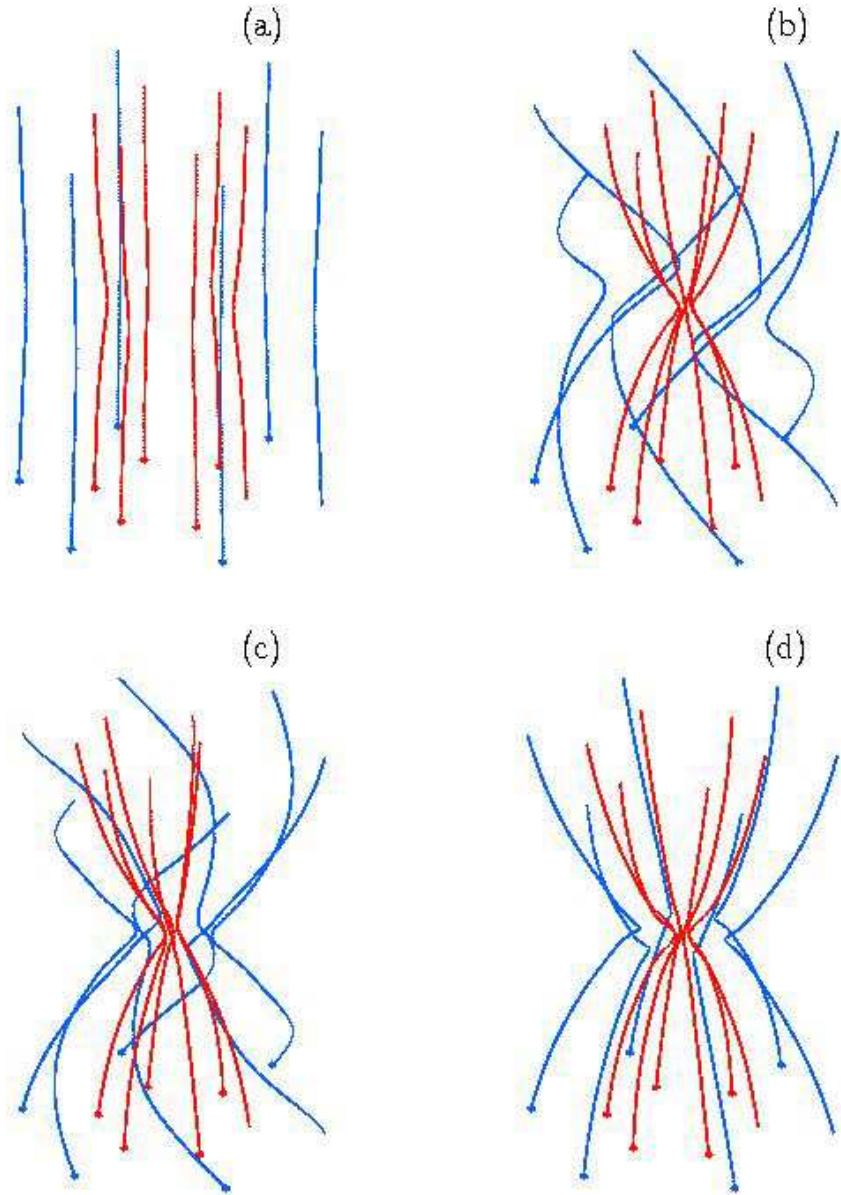}
\vspace*{-5.0cm}
\caption{These panels show the time
evolution of the magnetic field lines associated with a
Schwarzschild black hole and an accretion disk. The field lines
are traced from the bottom ($z=-8r_{\rm S}$) with the different
radius (red: $r = 2.67r_{\rm S}$; blue:
$r = 5.33r_{\rm S}$) at (a) $t=0.0\tau _{\rm S}$,
(b) $t=39.2\tau _{\rm S}$, (c) $t=60.0\tau _{\rm S}$, and (d)
$t=128.9\tau _{\rm S}$.} 
\label{blinefig}
\end{figure}

\begin{figure}
\vspace*{-6.5cm}
\plotone{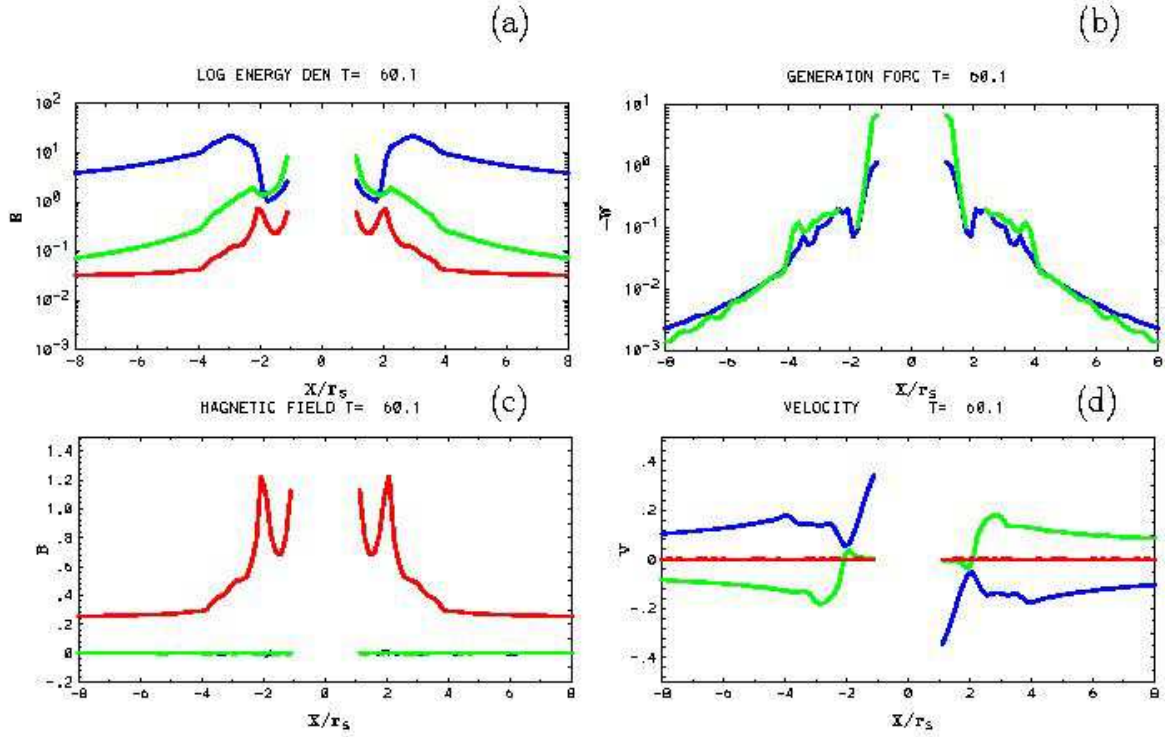}
\vspace*{-6.0cm}
\caption{This plot represents the one dimensional slice through
the simulation domain at $z = 0$ at  $t=60.0\tau _{\rm S}$. The
location of the black hole is excluded in the plots. Frame (a)
shows the proper mass density $\rho$ (blue), proper pressure $p$ (green),
and magnetic field energy $B_{\rm z}^{2}/2$ (red).
Frame (b) shows the power $-W_{\rm gp}$ (green)
and $-W_{\rm EM}$ (blue) associated with the gas pressure force and
electromagnetic force which contribute to
decelerating the accretion disk plasma.  Frame (c) shows the individual
magnetic field components $B_{\rm z}$ (red) and $B_{\rm y}$ (green),
and $B_{\rm x}$ (blue). Frame (d) shows the velocity components
$v_{\rm x}$ (blue), $v_{\rm z}$ (red), and $v_{\rm y}$ (green).}
\label{z0fig}
\end{figure}

\begin{figure}
\vspace*{-6.5cm}
\plotone{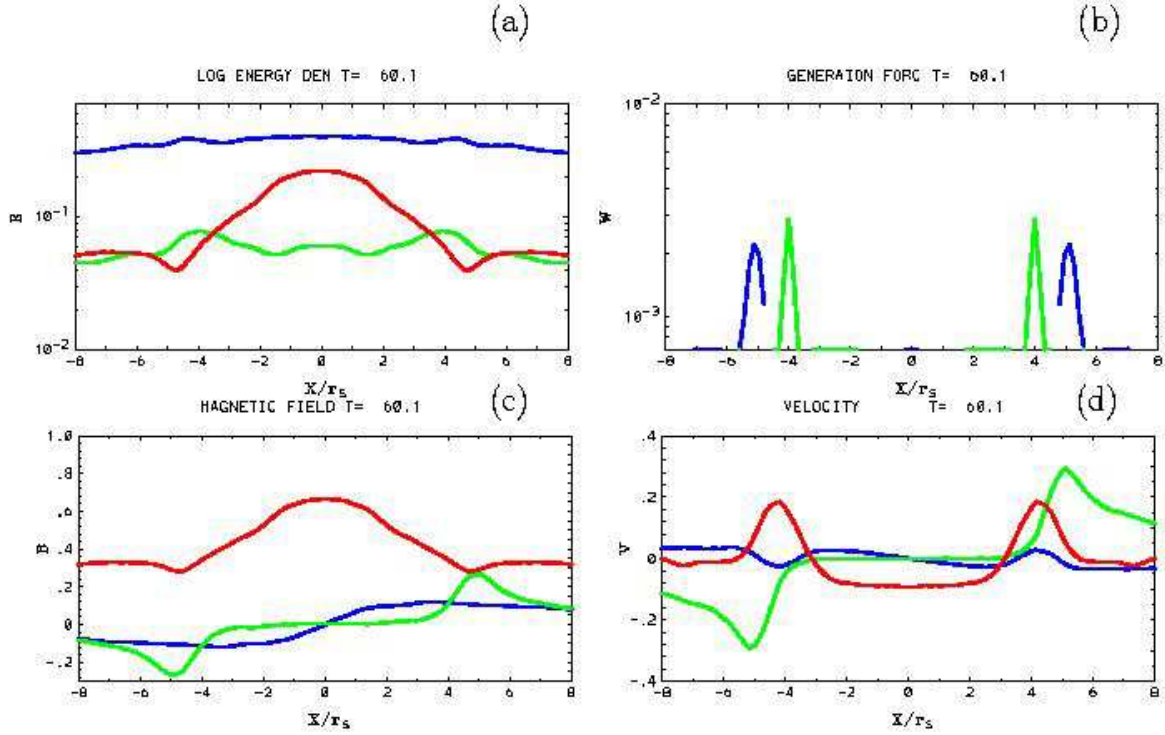}
\vspace*{-6.0cm}
\caption{This plot represents various physical quantities on the
$z = 5.6 r_{\rm S}$ surface at $t=60.0\tau _{\rm S}$.  Frame (a) shows
the values for proper mass density $\rho$ (blue), proper pressure $p$ (green)
and z component magnetic field energy  $B^{2}_{\rm z}/2$ (red).  Frame (b) is
the power contribution from the pressure, $W_{\rm gp}$ (green) and the
electromagnetic force, $W_{\rm EM}$ (blue).
We can find the two-layer acceleration region in the jet. Frame
(c) shows the individual magnetic field components $B_{\rm x}$
(blue), $B_{\rm y}$ (green), and $B_{\rm z}$ (red).  Frame (d) has
the components of velocity $v_{\rm x}$ (blue), $v_{\rm y}$
(green), and $v_{\rm z}$ (red).}
\label{z5fig}
\end{figure}

\begin{figure}[ht]
\vspace*{-6.5cm}
\plotone{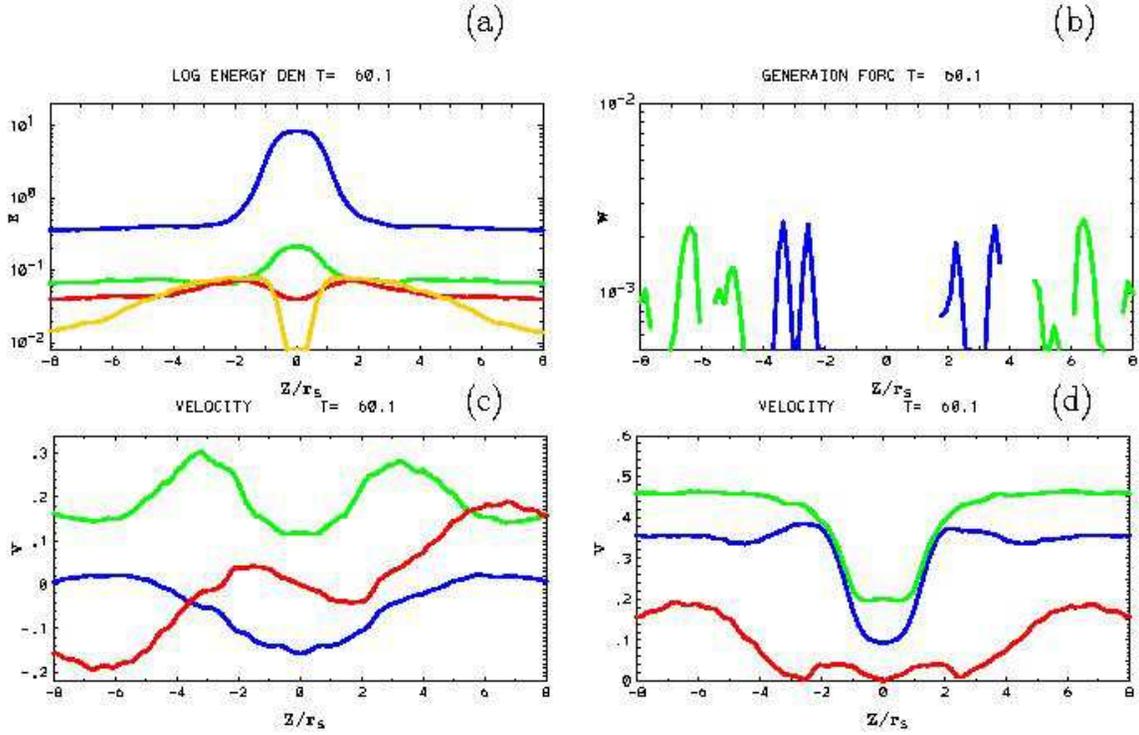}
\vspace*{-6.0cm} \caption{This plot represents various physical
quantities on the $x = 4.48r_{\rm S}$ surface at $t=60.0\tau _{\rm
S}$. Frame (a) shows the values for proper mass density $\rho$
(blue), proper pressure $p$ (green), z component magnetic field
energy $B^{2}_{\rm z}/2$ (red), and perpendicular magnetic field
energy $B^{2}_{\rm x} +B^{2}_{\rm y}$ (yellow). Frame (b) is the power
contribution of the pressure, $W_{\rm gp}$ (green) and the
electromagnetic force, $W_{\rm EM}$ (blue).
Frame (c) shows the components of the velocity $v_{\rm y}$ (green),
$v_{\rm z}$ (red), and $v_{\rm x}$ (blue). Frame (d) shows the
Alfv\'{e}n velocity $v_{\rm A}$ (blue), the sound velocity $v_{\rm S}$
(green), and the velocity $|v_{\rm z}|$ (red).}
\label{xfig}
\end{figure}

\begin{figure}[ht]
\vspace*{-6.5cm}
\plotone{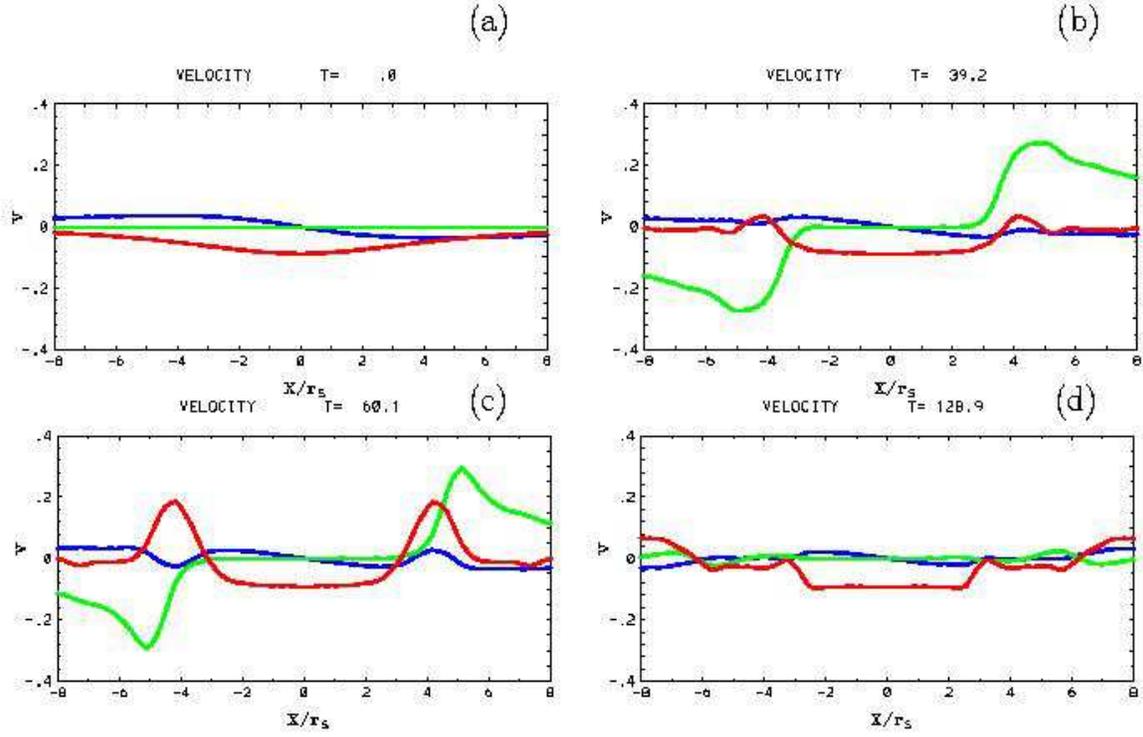}
\vspace*{-6.0cm}
\caption{The time evolution of one dimensional slice of velocities
through the simulation domain at $z = 5.6r_{\rm S}$ is plotted at (a)
$t=0.0\tau _{\rm S}$, (b) $t=39.2\tau _{\rm S}$, (c)
$t=60.0\tau _{\rm S}$, and (d) $t=128.9\tau _{\rm S}$. The blue, green,
and red curves present $v_{\rm x}, v_{\rm y}$, and $v_{\rm z}$,
respectively.}
\label{ztfig}
\end{figure}

\end{document}